\documentstyle[preprint,prd,aps]{revtex}
\tightenlines
\begin{document}

\title{ The Color Singlet Relativistic Correction in $J/\psi$ Photoproduction }
\par\vskip20pt
\author{C.B. Paranavitane\footnote{email: cbp@physics.unimelb.edu.au}
\and B.H.J. McKellar}
\date{\today}
\preprint{UM-P-98/55, RCHEP98/14}
\address{{\it Research Center for High Energy Physics,
School of Physics,\\ University of Melbourne, Parkville, Victoria 3010,
     Australia}}
\author{J.P. Ma\footnote{email: majp@itp.ac.cn}}
\address{ Institute of Theoretical Physics,
 Academia Sinica, \\P.O. Box 2735, Beijing 100080, China}

\vskip 30pt

\begin{center}
\maketitle
\end{center}
\begin{abstract}
The $O(v^2)$ relativistic correction for inelastic  $J/\psi$
photoproduction, in which heavy quark pairs are in the dominant Fock
state of the quarkonium, is studied in the framework of NRQCD
factorization. An assessment of its significance, particularly in
comparison to the color octet contributions, is made. It is found that the
impact on the energy distribution is negative in certain regions of phase
space. The predictions are compared with photoproduction data
from DESY-HERA. 

\end{abstract}

\def\s{\hat{s}}
\def\t{\hat{t}}
\def\uh{\hat u}
\def\st{(\s-4 m^2)}
\def\tt{(\t -4 m^2)}
\def\ts{(\s + \hat t)}
\def\u{(4 m^2-\ts)}
\def\bbbone{{\mathchoice {\rm 1\mskip-4mu l} {\rm 1\mskip-4mu l}
          {\rm 1\mskip-4.5mu l} {\rm 1\mskip-5mu l}}}
\newcommand{\um}{\bbbone}
\newcommand{\psid}{\psi^{\dagger}}
\newcommand{\chid}{\chi^{\dagger}}
\newcommand{\etad}{\eta^{\dagger}}
\newcommand{\zetad}{\zeta^{\dagger}}
\newcommand{\Db}{{\bf D}}

\newpage

\section{Introduction}
\label{intro}
Recent theoretical developments have been responsible for a resurgent
interest in quarkonia physics: a factorized form for quarkonia production
rates \cite{BFL}, based on the effective field theory, Non Relativistic
QCD (NRQCD) \cite{NRQCD}, has been proposed. In this formalism, the
inclusive
production rates can be expressed as a function of the rates for
creating pairs of free heavy quarks, $Q \bar{Q}$, in various
configurations,
and by a series of
non-perturbative matrix elements. The former may be computed using
perturbative QCD, while the matrix elements, which describe the
transition of the $Q \bar{Q}$ into a hadron, are defined in NRQCD.
Contrary to previous approaches, NRQCD factorization allows a
systematic determination of relativistic corrections, which describe
the effects of a heavy quark moving at a small velocity $v$ in the rest
frame of the quarkonium. The velocity scaling rules of NRQCD \cite{mag},
which reveal the power law behavior with respect to $v$ of the
operators of the effective field theory, may be used to organize the
matrix elements into a hierarchy. Although each quarkonium resonance is
identified by the $Q \bar{Q}$ having a particular set of quantum
numbers (such a $Q \bar{Q}$ state is the dominant Fock state of
the quarkonium), the relativistic corrections to the production rates
for that quarkonium, which accommodates the  relative motion of the $Q$ and 
$\bar Q$, may introduce a $Q \bar Q$ state with other quantum numbers.  
For example, the pair may be in a color octet configuration.
The corresponding channels in the production rate, despite being at
higher orders  in $v$, can be phenomenologically significant in certain
kinematic domains, due to the particulars of the production process.

One of the experiments used to test NRQCD factorization is
photoproduction of quarkonia, in which a real photon collides with a
nucleon. Charmonium photoproduction has been investigated in the 
framework of the color singlet model \cite{berger}, and more recently 
using NRQCD factorization \cite{KLS}, whose predictions have been
subject to a comparison with preliminary data from HERA \cite{H1,ZEUS}. In
addition to
the channel in which the $Q \bar{Q}$ is in the dominant Fock state of
the $J/\psi$, corrections, corresponding to the $Q
\bar{Q}$ in spin singlet and spin triplet, S and P wave, color octet
states, have been considered in the studies. The numerical 
values of the color octet matrix elements,  were estimated in an
earlier work \cite{CL}, by fitting theoretical predictions for prompt
charmonium production in $p\bar{p}$ processes, to  data from the
Fermilab-Tevatron. Using these values, however, leads to 
a significant discrepancy  with the photoproduction rates at HERA: 
the contributions from octet channels overwhelm the data in certain
regions of phase space. 

A number of uncertainties which enter into calculations may be
responsible for the inconsistency between the HERA and Tevatron data
for charmonium production, when viewed from the perspective of NRQCD.
Among these are uncertainties in parton distribution functions, the
value of the charm quark mass, higher twist effects \cite{Ma}, and
higher order perturbative  QCD corrections. In the work by Kniehl
\cite{Kniehl}, important higher order perturbative QCD effects, due to
initial state multiple gluon radiation, were included when analyzing
hadroproduction at the Tevatron. The re-estimated matrix elements have
values significantly smaller than the originally determined 
ones. Correspondingly, the HERA data is in much better agreement with
the NRQCD predictions. This is true throughout most of phase space, 
except near certain boundaries where the octet contributions diverge.
The authors in \cite{ben}, have suggested this is due to the  
non-relativistic expansion breaking down in such regions. They have 
proposed a resumation of the non-relativistic expansion, which requires
the introduction of universal structure functions. 

The contributions to $J/\psi$ photoproduction from the color octet
channels are at order $v^4$ in the small velocity expansion, relative
to the leading order, color singlet term. There is a relativistic  
correction at order $v^2$, incorporating the effects of  the relative
motion
of the heavy quark pair in the dominant Fock state of $J/\psi$.  
The purpose of this paper is to examine this correction in the framework
of
NRQCD factorization, which requires the introduction of an additional
matrix element. Previously, models have been constructed to
describe such an order $v^2$ effect. The one parameter model presented
in \cite{WKM} has been used to examine quarkonia photoproduction by Jung
et al. \cite{jung}; a two parameter model has also been proposed
\cite{khan}. The $O(v^2)$ contribution has not been computed
within NRQCD. 
It has been pointed out by Bodwin et al. \cite{BFL},
that 
corrections of $O(v^2)$ may be as important as radiative corrections of
$O(\alpha_s)$, which  have already been
calculated for  photoproduction \cite{NLO}, and found to be
significant. Without a detailed understanding of
the $O(v^2)$ term, it is not possible to make an assessment of the
significance of the color octet terms: the $O(v^2)$ contribution is
clearly necessary for consistency, according to the power counting
rules of NRQCD.  

The remainder of the paper is organized as follows. In the next section,
some details of the calculation of the relativistic correction is
described. In section \ref{num}, numerical results, indicating the
impact of the $v^2$ correction, are presented, together with a
comparison with  data from HERA.  Section \ref{end} contains some
concluding remarks.

\section{Photoproduction}
\label{photo}
NRQCD factorization allows the inclusive cross section for the
photoproduction of $J/\psi$ to be written in the following form:
\begin{equation}
\frac{d \hat{\sigma}}{d \hat{t}}
^{_{_{_{_{_{_{_{\gamma+i \rightarrow J/\psi +X}}}}}}}}=
\sum_{n} \frac{F_{n}}{m^{d_{n}-4}} 
\langle {\cal O}_{n}^{J/\psi} \rangle.
\label{xsection}
\end{equation}
The coefficients, $F_{n}$ depend on the kinematic invariants of the
process in which a photon $\gamma$ and a parton $i$ from the initial
hadron, react to produce a $Q \bar{Q}$. Here $m$ is the mass of heavy quark.  
The production is a short distance process, and is therefore calculable
using QCD  perturbation theory. The subscript $n= ^{2S+1} 
L_{J}^{(1,8)}$ specifies the spin $S$, orbital angular momentum $L$,
and the  total angular momentum $J$, of the $Q \bar{Q}$. The $1$ and
$8$ means that  the pair may be in a color singlet or color octet
state. The matrix elements $\langle {\cal O}_{n}^{J/\psi} \rangle$, of
mass dimension  $d_{n}$ describes the transition of the $Q \bar{Q}$
in the state $n$, into a $J/\psi$. Although  $J/\psi$ is identified as a
$Q\bar{Q}$
mainly in a $^{3} S _{1}^{(1)}$ configuration, the short distance
process may produce a heavy quark pair which is not in a $^{3} S
_{1}^{(1)}$ state. The formation of the bound state
can
involve the emission of soft partons,  which return the  pair into the
appropriate spectroscopic configuration. The relative importance of
various terms in the sum  in Eq.~(\ref{xsection}), may be determined by how
each term is suppressed by powers of the strong coupling constant,
$\alpha_{s}$, whose dependence lies in $F_n$, and the velocity of the
heavy quark in the rest frame of the quarkonium $v$, appearing in
$\langle {\cal O}_{n}^{J/\psi} \rangle$. For $J/\psi$, $v^2 \approx
0.3$.  

The dominant short distance contribution to inelastic photoproduction
of $J/\psi$ is the photon-gluon fusion process
\begin{eqnarray} 
   \gamma(k)  +  G(g_1)  \rightarrow 
Q \bar{Q}_n(P) +G(g_2).
\label{pro}  
\end{eqnarray}
The four momenta are specified in parenthesis. The process
(\ref{pro}) at leading order in QCD perturbation theory is represented
in Figs. 1(a) and  1(b). Both diagrams allow the heavy quark
pair to be produced in a color octet state, while the 
$Q\bar{Q}$ can only to be in a color singlet state in 
Fig.~1(a).
Subprocesses involving light quarks in the initial state are
considerably suppressed at the energies available at HERA \cite{KLS}.
Resolved
processes, in which the photon breaks up into partons, which
subsequently interacts with the nucleon, have been
analyzed within NRQCD factorization\cite{CK}. These are  important in
the kinematic regime $z \leq 0.3$, where $z=E_{J/\psi}/E_{\gamma}$,
in which the energies are defined in the nucleons rest frame. 

The formation of $J/\psi$ from a $Q\bar{Q}$  at leading order in the
small velocity expansion is described by the matrix element
\begin{equation}
\left\langle {\cal O}^{J/\psi}(^{3}S_{1}^{(1)})  \right\rangle = 
\left\langle 0 \left|
\chi^{\dagger} \sigma^{i} \psi
\left( \sum_{S,J_z}\left|J/\psi+S \left\rangle \right\langle J/\psi+S \right| \right)
        \psi^{\dagger} \sigma^{i} \chi
\right| 0 \right\rangle.
\label{lome} 
\end{equation}
The field $\psi$ annihilates a heavy quark, while $\chi$ creates a heavy 
anti-quark. The matrix element is proportional to the rate of 
production of a $J/\psi$ from the NRQCD vacuum, via an intermediate
state in which the heavy quark pair is in a $^{3} S_{1}^{(1)}$
configuration. $S$ represents light hadrons in the final state, and 
$J_z$ is the helicity of the $J/\psi$, both of  which are summed over. 

The relativistic correction, appearing at $O(v^2)$ relative to the leading
contribution, in which the heavy quark pair is also in the dominant Fock
state of the $J/\psi$, involves the matrix element
\begin{equation}
\left\langle  {\cal P}^{J/\psi}(^{3}S_{1}^{(1)})  \right\rangle = 
\left\langle 0 \left|
\frac{1}{2}\left[\chi^{\dagger} \sigma^{i} \psi
\left(\sum_{S}|J/\psi+S \rangle \langle J/\psi+S |\right)
        \psi^{\dagger} \sigma^{i} \left(-\frac{i}{2}
\stackrel{\leftrightarrow}{\bf{D}}\right)^{2} \chi
+ h.c\right]
\right| 0 \right\rangle,
\label{eq:correction}
\end{equation}
where $\chid \stackrel{\leftrightarrow}{\bf{D}} \psi =
\chid (\Db) \psi - (\Db \chi)^{\dagger} \psi$.
Therefore, the differential cross section  
to $O(v^2)$ takes the form,
\begin{eqnarray}
\frac{d \hat{\sigma}}{d \hat{t}}
^{_{_{_{_{_{_{_{\gamma+G \rightarrow J/\psi +X}}}}}}}}=
\frac{F(^{3} S_{1}^{(1)})}{m^{2}}
 \langle {\cal O}^{J/\psi}(^{3} S_{1}^{(1)}) \rangle +
\frac{G(^{3} S_{1}^{(1)})}{m^{4}}
 \langle {\cal P}^{J/\psi}(^{3} S_{1}^{(1)}) \rangle  +O(v^4).
\label{dxs} 
\end{eqnarray}
Of the octet contributions which appear at $O(v^4)$,
the $^{1} S_{0}^{(8)}$ and $^{3} P_{J}^{(8)}$ channels were found to be
the most 
important. The structure of the matrix elements which describe these 
transitions, are similar to Eqs.(\ref{lome}) and (\ref{eq:correction}),
except for
the operators between the quark and anti-quark fields; they may be found in
the
literature\cite{BFL}. The aim in this 
section is to determine the coefficient $G(^3 S_1^{(1)})$, containing
the short distance information associated with the $v^2$ correction.   

The low energy equivalence of full QCD and NRQCD allow the short distance
coefficients $F_n$  to be determined by the following matching 
condition:   
\begin{equation}
\left. 
\frac{d \hat{\sigma}}{d \hat t}
^{_{_{_{_{_{_{_{\gamma+G \rightarrow Q \bar{Q}_n +X}}}}}}}}
\right|_{pQCD} = 
\sum_n \left. \frac{F_{n}}{m^{d_{n}-4}}
\langle 0 | {\cal O}_{n}^{Q \bar{Q} } | 0 \rangle
\right|_{pNRQCD}. 
\label{match} 
\end{equation}
On the left hand side of Eq. \ref{match}), the cross section for the 
production of a free $Q \bar{Q}$ is computed using perturbative QCD. 
The matrix elements on the right hand side are evaluated using 
perturbative NRQCD. Both sides are expanded in powers of $\bf q$, the 
momentum of the heavy quark in the rest frame of the quarkonium. The
$F_n$ are then found by comparing powers of $\bf q$. The QCD side of
the matching condition involves expanding the amplitude for the process,
Eq. (\ref{pro}), in powers of ${\bf q}$. The  leading terms contain $Q
\bar{Q}$'s in the dominant Fock state of the $J/\psi$, $^3 S_{1}^{(1)}$,
as well as some octet configurations: $^3 S_{1}^{(8)}$ and $^1
S_{0}^{(8)}$. Terms at order ${\bf q}$ in the  amplitude project out
heavy quark pairs in $^3 P_J^{(8)}$ states, but do not contain a heavy quark
pair 
in $^3 S_{1}^{(1)}$ states. In order to determine the $O(v^2)$
contribution in (\ref{dxs}), it is necessary to expand the QCD
amplitude  to $O({\bf q}^2)$: at this order, heavy quark pairs in $^3
S_{1}^{(1)}$ configurations are present. The reason that the color
octet contributions appear at a higher order in $v$, in (\ref{dxs}),
despite being at a lower order in the perturbative amplitude, is
because the power counting is different for hadronic states: additional
$v$ suppression appears in the hadronic projection operators of the
color octet matrix elements. 

Some conventions used in \cite{BC} will be adopted in this work. The
four momenta of the heavy quark and anti-quark is
\begin{eqnarray}
p^{\mu} = \frac{P^{\mu}}{2} + \Lambda^{\mu}_{i} q^{i}, \nonumber \\
\bar{p}^{\mu} = \frac{P^{\mu}}{2} - \Lambda^{\mu}_{i} q^{i},
\end{eqnarray}
where $\Lambda$ is the Lorenz boost matrix from the rest frame of the
$J/\psi$, to the frame in which it is moving with four momentum 
$P = (\sqrt{4(m^2+\vert {\bf q}\vert ^2)+{\bf P}^2},{\bf P})$.
The partonic Mandelstam invariants are defined in the standard manner: 
$\hat{s} = (k+g_{1})^2=(P+g_{2})^2$, $\hat{t} = (k-P)^2= (g_1-g_2)^2$ 
and $\hat{u} = (g_1-P)^2=(k-g_2)^2$. Only two of these  invariants are
independent, their relationship is given by
\begin{equation}
\hat{s}+\hat{t}+\hat{u} = P^{2} = 4 (m^{2}+\vert {\bf q}\vert^{2}).
\label{stu}
\end{equation}
We choose $\hat{u}$ to be expressed in terms of the other two
invariants, and $\vert {\bf q}\vert^2$, which is associated with   
$\langle {\cal P}^{J/\psi}(^{3} S_{1}^{(1)}) \rangle$. Two contraction
identities found in reference\cite{BC} will be frequently used in the 
calculation: 
\begin{eqnarray} 
g_{\mu \nu} \Lambda^{\mu}_{i} \Lambda^{\nu}_{j} & = & -\delta^{ij}
\nonumber, \\
\Lambda^{\mu}_{i} \Lambda^{\nu}_{i} & = & -g_{\mu \nu}+ 
\frac{P^{\mu} P^{\nu} }{P^{2}}.
\label{contractid}
\end{eqnarray}

First we will consider the QCD side of Eq. (\ref{match}). The transition
matrix for the process may be written,
\begin{eqnarray*}
{\cal T}_{\gamma+G \rightarrow Q \bar{Q}_n+G} =
\varepsilon_{\mu}(k) \varepsilon_{\alpha}^{a}(g_1) 
\varepsilon^{\ast b}_{\beta}(g_2) 
\bar{u}(p) T^{\mu \alpha \beta}_{a b} v(\bar{p}).
\end{eqnarray*}
The color indices of the gluons are specified by $a$ and $b$. Following
the
techniques used in \cite{BC}, $T^{\mu \alpha \beta}_{a b}$ is
decomposed in terms of the sixteen basis matrices $\bbbone,\gamma_{5},
\gamma_{\mu},\gamma_{\mu} \gamma_{5}$, and $\sigma_{\mu \nu}$. When
these are inserted between the quark and anti-quark spinors
$\bar{u}(p)$ and $v(\bar{p})$, ${\cal T}$ may be expressed in terms of
the color triplet, Pauli spinors, $\xi$ and $\eta$. Projecting out the
color singlet channel and expanding to $O(|\bf{q}|^{3})$, the transition
matrix takes the form
\begin{equation}
{\cal T} =
\varepsilon_{\mu}(k) \varepsilon_{\alpha}^{a}(g_1)
\varepsilon^{\ast b}_{\beta}(g_2) 
\frac{\delta_{ab}}{2 N_{c}}
(a^{\mu \alpha \beta}_{i} + b^{\mu \alpha \beta}_{i m n} 
q^{m} q^{n}) \xi^{\dag} \sigma^{i} \eta + ...
\end{equation}
The terms $a^{\mu \alpha \beta}_{i}$ and $b^{\mu \alpha \beta}_{i m n}$
are functions of the kinematic variables and the Lorenz boost matrices.
The ellipsis represents contributions in which the $Q \bar{Q}$ pairs are in 
color singlet $P$ wave states, spin singlet states, or terms which
contribute to processes occurring at $O(v^{3})$ or higher.
To obtain the rate, ${\cal T}$ is multiplied by it's complex conjugate, 
and
the appropriate sums and averages over spin and color degrees of freedom
are
made. Before the momentum space integration is carried out, a change of
variables is performed: $(p, \bar{p}) \rightarrow (P,q)$, to facilitate
the
matching as defined in Eq. (\ref{match}). The perturbative cross section
becomes
\begin{equation}
\frac{d \hat{\sigma}}{d \hat t}
^{_{_{_{_{_{_{_{\gamma+G \rightarrow Q \bar{Q}_n +X}}}}}}}} = 
-\frac{1}{16 \pi \hat{s}^2}
\left(\frac{1}{2^2.4 N_{c}^2}\right) 
\int \frac{d^3 {\bf q}}{(2 \pi)^3}
\left[ A_{i j} + \left(b_{ij}^{(2)}-\frac{A_{ij}}{2 m^2}\right) 
 \vert{\bf q}\vert^2  
+c^{(2)}_{i j m n} q^{m} q^{n} \right] 
\sigma^{i} \otimes \sigma^{j}  + ...
\label{lorcon}
\end{equation}
For convenience, the notation  suppressing the Pauli spinors and the sums 
over the spins has been employed\cite{BFL}. The coefficients appearing
in the transition matrix are related to the  ones appearing in
(\ref{lorcon}). Using the identities in (\ref{contractid}), the
contraction of the Lorenz indices relates the $A_{ij}$,
$b_{ij}^{(2)}$ and $c^{(2)}_{i j m n}$, to $a^{\mu \alpha \beta}_{i}$ and
$b^{\mu \alpha \beta}_{imn}$. 
\begin{eqnarray*}
a^{\mu \alpha \beta}_{i} a_{\mu \alpha \beta,j}^{\ast} &=&
A_{ij} + b_{ij}^{(2)} \vert{\bf q}\vert^2 + O(\vert {\bf q}\vert^3) \\
a^{\mu \alpha \beta}_{i} b_{\mu \alpha \beta,imn}^{\ast} +
b^{\mu \alpha \beta}_{imn} a_{\mu \alpha \beta,j}^{\ast}
& = &  c^{(2)}_{i j m n} + O(|\bf{q}|) 	
\end{eqnarray*}
It must be kept in mind that the factor $A_{i j}$ implicitly retains
some $\bf q$ dependence which will become fully exposed once
contractions over the remaining indices are carried out.
The term $A_{ij}/2 m^2$ is a consequence of the change of variables.
Using rotational symmetry, 
Eq. (\ref{lorcon}) takes the form
\begin{eqnarray}
\frac{d \hat{\sigma}}{d \hat{t}}
^{_{_{_{_{_{_{_{\gamma+G \rightarrow Q \bar{Q}_n +X}}}}}}}} & = &
-\frac{1}{16 \pi \hat{s}^2}
\left(\frac{1}{2^2.4 N_{c}^2}\right)
\int \frac{d^3 {\bf q}}{(2 \pi)^3}
\left[
\frac{1}{3} c_{ii}^{(0)} \sigma^j \otimes \sigma^j
+ \frac{1}{3} \left(3 c_{ii}^{(2)} +  b_{ii}^{(2)}-
\frac{c_{ii}^{(0)}}{2 m^2}\right)  \right.  \label{contract} \\ 
\sigma^j \otimes \sigma^j \vert{\bf q}\vert^2 \nonumber  
 & & \left. + \frac{1}{30} \left[ 
(4 c_{iijj}^{(2)}-c_{ijij}^{(2)}-c_{ijji}^{(2)}) 
 \vert{\bf q}\vert^2 \sigma^{m} \otimes \sigma^{m}+
(-2 c_{iijj}^{(2)} + 3 c_{ijij}^{(2)} + 3 c_{ijji}^{(2)}) 
{\bf q}.{\bf \sigma} \otimes {\bf q}.{\bf \sigma} 
 \nonumber \right] \right] .
\end{eqnarray}
Only the leading order terms in $|\bf{q}|^2$ have been kept in the
contractions of  $c_{ijmn}^{(2)}$ and $b_{ij}^{(2)}$. $c_{ii}^{(0)}$
and $c_{ii}^{(2)}$ are defined to be 
\begin{eqnarray*}
A_{ii}= c_{ii}^{(0)} + c_{ii}^{(2)}\vert{\bf q}\vert^2 .
\end{eqnarray*}
Equation (\ref{contract}) displays all the ${\bf q}^2$ dependence 
necessary to determine $G(^3 S_1^{(1)})$. The coefficients have been
expressed in terms of the invariants $\hat{s}$ and $\hat{t}$ using the
contraction identities. They are somewhat lengthy and will not be
presented here. 

Next the NRQCD side of the matching condition is considered. 
The $Q \bar{Q}$ differential cross section, only including the
contributions we are interested in, may be written
\begin{eqnarray} 
& & \frac{F(^{3} S_{1}^{(1)})}{m^{2}}
 \langle {\cal O}^{Q \bar{Q}}(^{3} S_{1}^{(1)}) \rangle +
\frac{G(^{3} S_{1}^{(1)})}{m^{4}}
 \langle {\cal P}^{Q \bar{Q}}(^{3} S_{1}^{(1)}) \rangle  + \nonumber \\
 &&
\frac{F(^{3} S_{1}^{(1)},^{3} D_{1}^{(1)})}{m^{4}}
 \langle {\cal P}^{Q \bar{Q}}(^{3} S_{1}^{(1)},^{3} D_{1}^{(1)})
\rangle+\cdots 
\end{eqnarray}
It is necessary to take into account another color singlet matrix 
element with $d_{n} = 8$,
\begin{eqnarray}
\left\langle {\cal P}^{Q \bar{Q}}(^{3} S_{1}^{(1)},^{3} D_{1}^{(1)}) 
\right\rangle =
\left\langle 0 \left| \frac{1}{2} \left[
\chi^{\dagger} \sigma^{i} \psi
\left(\sum_{S}|Q \bar{Q} \rangle \langle Q \bar{Q} |\right)
\psi^{\dagger} \sigma^{j} \left(-\frac{i}{2}\right)^{2} 
\stackrel{\leftrightarrow}{D}^{(i} \stackrel{\leftrightarrow}{D}^{j)}
\chi
+ h.c \right] \right| 0 \right\rangle.
\end{eqnarray} 
The symmetric traceless tensor, formed from the covariant 
derivatives is specified by
$ \stackrel{\leftrightarrow}{D}^{(i} \stackrel{\leftrightarrow}{D}^{j)} =
(\stackrel{\leftrightarrow}{D}^{i} \stackrel{\leftrightarrow}{D}^{j}+
\stackrel{\leftrightarrow}{D}^{j} \stackrel{\leftrightarrow}{D}^{i})/2 -
\stackrel{\leftrightarrow}{\bf D}.\stackrel{\leftrightarrow}{\bf D}/3$.
Here $\langle {\cal P}^{Q \bar{Q}}(^{3} S_{1},^{3} D_{1})\rangle$
describes the overlap of the transition amplitudes in which the $Q
\bar{Q}$ is in spin triplet $S$ and $D$ wave states. Therefore, $F(^{3}
S_{1},^{3} D_{1}) $ mixes with $F(^{3} S_{1})$ in perturbation theory.
$\langle {\cal P}^{Q \bar{Q}}(^{3}
S_{1},^{3} D_{1})\rangle$ appears at $O(v^{2})$
in perturbation theory, where quarks and gluons are
free. However, because the velocity counting is different for hadronic
states, the corresponding hadronic matrix element
$\langle {\cal P}^{J/\psi}(^{3}  S_{1},^{3} D_{1})\rangle$ is at order
$v^4$. This is because
it requires two chromo-electric dipole transitions to take a $Q
\bar{Q}$ in a $^{3} S_{1}^{(1)}$ configuration, to a $^{3}D^{(1)}_{1}$
state, costing a factor of $v^2$ in the projection operator of $\langle
{\cal P}^{J/\psi}(^{3} S_{1}^{(1)},^{3} D_{1}^{(1)}) \rangle$. The
covariant derivatives in the operator contributes another $v^2$.

Using a non-relativistic normalization convention, it is possible to
realize the right hand side of (\ref{match}) at leading order 
in perturbative NRQCD: 
\begin{eqnarray} 
& & \int \frac{d^3 {\bf q}}{(2 \pi)^3} \left[ 
\frac{F(^{3} S_{1}^{(1)})}{m^{2}} \sigma^{i} \otimes \sigma^{j} 
 + \frac{1}{3 m^{4}} \left( 3 F(^{3} S_{1}^{(1)},^{3} D_{1}^{(1)}) 
{\bf q} .{\bf \sigma} \otimes {\bf q}.{\bf \sigma} \right.
\right. \nonumber \\
& & \left. \left. \mbox{} +  3  \left[ G(^{3} S_{1}^{(1)})-   
 F(^{3} S_{1}^{(1)},^{3} D_{1}^{(1)}) \right] 
{\bf q}^2 \sigma^{i} \otimes \sigma^{i}  \right) \right].
\label{pNRQCD}
\end{eqnarray}
One may worry about the use of different normalization of states on 
either side of (\ref{match}). However this is not a
problem as only physical quantities are being considered, and  the Pauli
spinors, which are not shown explicitly, have the same normalizations: 
$ \xi^{\dagger} \xi = 1$ and $\eta^{\dagger} \eta = 1$. Braaten
et al. \cite{BC} defined the matrix elements  with  relativistic
normalizations, which were subsequently related to to the matrix elements
defined in NRQCD. 

By comparing Eq. (\ref{pNRQCD}) with Eq. (\ref{contract}), 
we finally obtain the short distance coefficients. The leading order color
singlet  
contribution is well known:
\begin{eqnarray}
 \frac{F(^{3} S_{1}^{(1)})}{m^2}  & =& 
         {\frac{1}{16 \pi \s^{2}}} \frac{32 m}{27}
(4 \pi \alpha_{s})^2  (4 \pi \alpha) e_{c}^{2}   \nonumber \\ 
  & &  \cdot \frac{\s^2 \st^2 + \t^2 \tt^2 + 
    (4 m^2 -\s-\t)^2 \ts^2}{\st^2 \tt^2 \ts^2}  .
\label{FS}
\end{eqnarray} 
The strong coupling constant and the fine structure constant are given by 
$\alpha_{s}$ and $\alpha$. The electromagnetic charge of the charm quark
in units
of $e$, is  $e_{c}$. As a consequence of Eq. (\ref{stu}), Eq. (\ref{FS}) is
expressed in terms of the combination $4 m^2-\hat{s}-\hat{t}$, instead
of $\hat{u}$. The perturbative coefficient of the relativistic correction for 
the color singlet channel is found to be
\begin{eqnarray}
\frac{G(^{3} S_{1}^{(1)})}{m^4} & = 
& \frac{1}{16 \pi \hat s^{2}} 32
(4 \pi \alpha)^2 (4 \pi \alpha) e_{c}^{2}
     (-6144 m^{10} {\s}^{2} + 1280 m^{8} {\s}{^3} + 960 m^{6} {\s}^{4}
 \nonumber \\ &&{} - 
   336 m^{4} {\s}^{5} + 28 m^{2} {\s}^{6}   -  4096 m^{10} {\s} {\t}
  - 512 m^{8} {\s}^{2} {\t} + 3200 m^{6} {\s}^{3} {\t} \nonumber\\ &&{}
 - 1008 m^{4} {\s}^{4} {\t
} +
   72 m^{2} {\s}^{5} {\t}  + {\s}^{6} {\t} - 6144 m^{10} {\t}^{2} -
   512 m^{8} {\s} {\t}^{2} + 3200 m^{6} {\s}^{2} {\t}^{2} \nonumber\\ &&{} -
   1408 m^{4} {\s}^{3} {\t}^{2}  + 92 m^{2} {\s}^{4} {\t}^{2} + 3 {\s}^{5
} {\t}^{2} +
   1280 m^{8} {\t}^{3}
 + 3200 m^{6} {\s} {\t}^{3}\nonumber \\ &&{} - 1408 m^{4} {\s}^{2} {\t}^{3}  +
   112 m^{2} {\s}^{3} {\t}^{3} + 5 {\s}^{4} {\t}^{3} + 960 m^{6} {\t}^{4} -
   1008 m^{4} {\s} {\t}^{4}
   \nonumber\\ &&{} + 92 m^{2} {\s}^{2} {\t}^{4} + 5 {\s}^{3} {\t
}^{4} -
   336 m^{4} {\t}^{5} + 72 m^{2} {\s} {\t}^{5} + 3 {\s}^{2} {\t}^{5} +
   28 m^{2} {\t}^{6}
 + {\s} {\t}^{6})/ \nonumber\\ &&{} 
\; \;\; \; \,  (81 m ({\s}-4 m^{2})^{3}({\t}-4 m^{2})^{3}({\s}+{\t})
^{3}).
\label{relcor}
\end{eqnarray}
 
This result is different to the one found in  \cite{jung}. First, 
the relativistic correction in this work is expressed in terms of a 
matrix element defined in NRQCD; in \cite{jung}, it is 
correspondingly  a parameter $\epsilon$, defined via 
\begin{equation} 
 M_{J/\psi}=2m+\epsilon,
\end{equation} 
where $\epsilon$ was taken to be positive. As pointed out in \cite{Ma}, 
this parameter can be negative and can be defined as a matrix element
in NRQCD which is different to the one appearing in Eq. (5). 
Secondly, the correction calculated in \cite{jung}, is based on the model
proposed in \cite{WKM}, in which the contribution of the momentum ${\bf
q}$ to the energy of the heavy quark is neglected-that is the energy is
taken to be $M_{J/\psi}/2$, in the quarkonium rest frame; it really
should be $\sqrt{m^2+{\bf q}^2}$. This implies that our result
cannot be obtained from that presented in \cite{jung} by setting
$M_{J/\psi}=2 m$, and the parameter $\epsilon$ proportional to
$\langle {\cal P}^{H/\psi}(^{3} S_{1}^{(1)}) \rangle$.

\section{Numerical Results}
\label{num}
In this section an assessment of the  significance of the order $v^2$
correction shall be made, using the analytic result (\ref{relcor}). The
predictions will be compared to photoproduction  data from HERA. In
\cite{KLS}, the color octet channels $^1S_0^{(8)}$ and $^3P_J^{(8)}$ were
found to be important in photoproduction of $J/\psi$. The corresponding
matrix elements were taken to have numerical values consistent with the
linear combination \cite{CL}
\begin{equation}
\langle {\cal O}^{J/\psi}(^1 S_0^{(8)}) \rangle +
\frac{3}{m^2} \langle {\cal O}^{J/\psi}(^3 P_0^{(8)}) \rangle = 6.6
\times 10^{-2} \ {\rm GeV}^3,
\label{SplusPorig}
\end{equation}
determined in the original analysis of hadroproduction rates at the
Tevatron. However, these values
lead to predictions which are significantly greater than the
experimental results. The inclusion of effects due to initial state
gluon radiation modifies the linear combination of the matrix elements
thus~\cite{Kniehl}:
\begin{equation}
\langle {\cal O}^{J/\psi}(^1 S_0^{(8)}) \rangle +
\frac{3.54}{m^2} \langle {\cal O}^{J/\psi}(^3 P_0^{(8)}) \rangle = 5.27
\times 10^{-3} \ {\rm GeV}^3.
\end{equation}
In this
study, we shall
assume the latter, and take $\langle {\cal O}^{J/\psi}(^1
S_0^{(8)}) \rangle = \langle {\cal O}^{J/\psi}(^3 P_0^{(8)}) \rangle/m^2
= 1.2 \times 10^{-3} \ {\rm GeV}^3$.  
The leading order color singlet matrix element is taken to have the value:
\begin{equation}
\langle {\cal O}^{J/\psi}(^3 S_1^{(1)}) \rangle = 1.3 \ {GeV}^3. 
\end{equation}
A naive estimate for $\langle {\cal P}^{J/\psi}(^3 S_1^{(1)}) \rangle$
shall be
used, based on the power law suppression of $\langle {\cal P}^{J/\psi}(^3
S_1^{(1)}) \rangle$ relative to 
$\langle {\cal O}^{J/\psi}(^3 S_1^{(1)}) \rangle$: 
\begin{equation}
\langle {\cal P}^{J/\psi}(^3 S_1^{(1)}) \rangle = 0.85 \ {\rm GeV}^5.
\end{equation} 
The other inputs for the computations are as follows: the strong
coupling constant at the charm mass scale is taken to be
$\alpha_{s}(m)=0.3$, with $m=1.48$ GeV. The MRSA determination of the 
gluon structure function is used \cite{MRSA}, with  
$\Lambda_{QCD}^{\bar{MS},n_f=4}=300$ MeV, and $Q^2=(2m)^2$. 

The total cross section as a function of the center of mass energy is
considered first. The constraints: $z \leq 0.8$ and $p_t \geq 1$ 
GeV, valid at HERA, are adopted. The K factor, consistent with these cuts,
from the NLO corrections to the color singlet channel \cite{NLO} is $K
\approx 1.8$.
Figure 2 shows the total cross section as a function of the
photon-nucleon center of mass energy, $\sqrt{s}$.
The leading order color singlet contribution (long dashed line), accounts
quite well for most of the data from HI \cite{Beate} (circles) and ZEUS
\cite{ZEUS} (squares). The $v^2$ correction (short dashed line), together
with the color octet terms (dot-dash line), contributes 
only marginally.  The octet contributions are greater 
than the $O(v^2)$ correction by approximately a factor of four, for large
center of mass energies, with the values of the matrix elements that have
been adopted. If the octet matrix elements were
chosen to be consistent with (\ref{SplusPorig}), then the $v^2$  
correction would  be smaller by approximately a factor of ten,
relative
to the octet contributions.   
However, $\langle {\cal O}^{J/\psi}(^3 S_1^{(1)}) \rangle$ may be
somewhat larger than $0.85 \ {\rm GeV}^5$,
therefore both corrections  may have similar magnitudes.

Next, the $d \sigma /dz$ distribution is considered, as a function of $z$.
Figure 3(a) shows the ratio of the $v^2$ to the octet contributions, at
$\sqrt{s}=100$ GeV (solid line) and $\sqrt{s}=14.7$ GeV (dashed line). At
$\sqrt{s}=100$ GeV the cut,  $p_t \geq 1$ GeV, consistent with
results from  HERA is adopted, while at $\sqrt{s}=14.7$, $p_t \geq 0.1$
GeV, valid for  the European Muon Collaboration (EMC). 
For  $\sqrt{s}=100$ GeV, it can
be seen that the the $v^2$ channel dominates at low $z$. For $z >
0.63$ the ratio is negative, indicating that the $v^2$ distribution is
negative in that region. At $z \approx 0.7$ the ratio has a minimum value
of about $0.45$, meaning that in this region, the $v^2$ contribution is
significant with respect to the color octet channels. Again, if the
original determinations of the octet matrix elements were adopted, the
$v^2$ contributions will give a negligible result at intermediate and
large values of $z$, while at low $z$, they would be comparable to the
color octet contributions. 
At the EMC energy regime, $\sqrt{s}=14.7$ GeV, $v^2$ corrections are
relatively more substantial at low $z$, while at larger values, not as
important. Figure 3(b) is the $z$ distribution at the center
of mass energy $\sqrt{s}=100$ GeV.
The long dashed line represents the leading order contribution. The sum of
the leading order and $v^2$ correction are shown with a dot-dash; the
sum of the leading order and octet contributions is shown with the dashed
line. The net result is shown with a solid line. The circles represent
data from the H1 Collaboration  \cite{Beate}.
The $v^2$ correction
tends to decrease the distribution for larger values of $z$, while at
smaller values, it is enhanced. Figure 3(c) shows the
distribution at $\sqrt{s}=14.7$ GeV, displaying a similar
behavior, although the $v^2$ effect is not as pronounced. The data is from
the EMC  \cite{emc}.
It should be noted that the large $z$ region where the $v^2$
contribution is most substantial, is also where the non relativistic
expansion breaks down \cite{ben}. Our analysis,
leads to a qualitatively different behavior for the energy distribution,
than in the work by Jung et al. \cite{jung}. In their case there is an
enhancement of  $d \sigma /dz$ for large values of $z$.

The transverse momentum distribution for photoproduction was also
examined. However, it was found that the $v^2$ correction leads to a
negligible contribution to the overall differential rate. It is expected
that a similar $v^2$ correction appears in $p+ {\bar p} \rightarrow J/
\psi +X$ processes at the Tevatron, where color octet fragmentation
channels dominate at large transverse momentum. The $v^2$ correction can
be
obtained by realizing that it originates via a process similar to that
depicted in figure 1(a), except with the initial photon replaced by a
gluon. The color singlet rates, including the $v^2$ correction is obtained
by multiplying $F(^{3} S_{1}^{(1)})$ and $G(^{3} S_{1}^{(1)})$ by the
factor $5 \alpha_{s}/(96 \alpha e_{c}^{2})$, which takes into account the
differences in the initial state. However it was found that the $v^2$
correction contributes insignificantly to the transverse momentum
distribution for moderate transverse momenta, 
at the Tevatron center of mass energy $\sqrt{s} = 1.8$ TeV.

\section{Conclusion}
\label{end}
In this work, the relativistic correction to the color singlet channel in
$J/\psi$ photoproduction was calculated using the NRQCD factorization
formalism.
The correction is described by an additional matrix element which appears at
$O(v^2)$ relative to the leading order color singlet term, according to the
velocity scaling rules of NRQCD. This work contrasts with previous studies 
in that it  draws qualitatively different conclusions about the 
behavior of observable. Specifically, the contribution leads to a decrease
in the energy distribution for large values of $z$, while for smaller values
there is an enhancement of the differential cross section. The effect of the 
correction in photoproduction seems to be more significant for larger center 
of mass energies. In some regions of phase space, the $v^2$ correction is
comparable to the color octet contributions, with the values of the
parameters that were used. However, it has been noted earlier that the
rates are sensitive to $\alpha_s(m)$, $m$, and the choice of parton
distribution functions, which are not precisely known. Certainly it seems
that the uncertainty in these parameters, correspond to variations in the
rates which are of similar magnitude, or greater than the effect of the
$O(v^2)$ correction \cite{NLO}. Therefore, at the present level of
precision, it is not possible to observe the effect of the correction
in experiments.   

{\bf Acknowledgments}
This work is supported by the Australian Research Council. 

\newpage
\centerline{\bf Figure Captions}
\par
\noindent
Figure 1: (a) One of six Feynman diagrams for the production of  a       
$Q\bar Q$ pair. The others are obtained through permutations 
of photon and gluon lines. The $Q\bar{Q}$ may be in a color singlet 
or color octet configuration.  
(b) A typical diagram in which the $Q\bar{Q}$'s are only in color octet 
states.  

Figure 2: The total cross section $\sigma (\sqrt{s})$ (nb) as a
function of the center of mass energy $\sqrt{s}$ (GeV). The leading
order color singlet contribution is shown with a long dashed line. The
$O(v^2)$ correction is the dot-dash line; the octet contribution is the
short dash line. The total rate is the solid line.

Figure 3: (a)  $d \sigma/dz$: the ratio of the $v^2$ contribution to the
octet contributions as a function of $z$. The solid line is at
$\sqrt{s}=100$ GeV, and the dashed line is at $\sqrt{s}=14.7$ GeV.
(b) $d \sigma/dz$ (nb) as a function of z, at the center of mass
energy $\sqrt{s}=100$ GeV (HERA). The long dashed line is the leading
order color singlet contribution. The sum of the leading order and
$O(v^2)$ channels is the  dot-dash line. The sum of the leading order and
octet contribution is the dashed line. The solid line represents the total
rate. 
The circles are the data from the H1 collaboration.
(c)  $d \sigma/dz$ (nb) at $\sqrt{s}=14.7$ GeV. The lines represents
contributions as described in (b). Data from EMC are shown with
solid circles. 

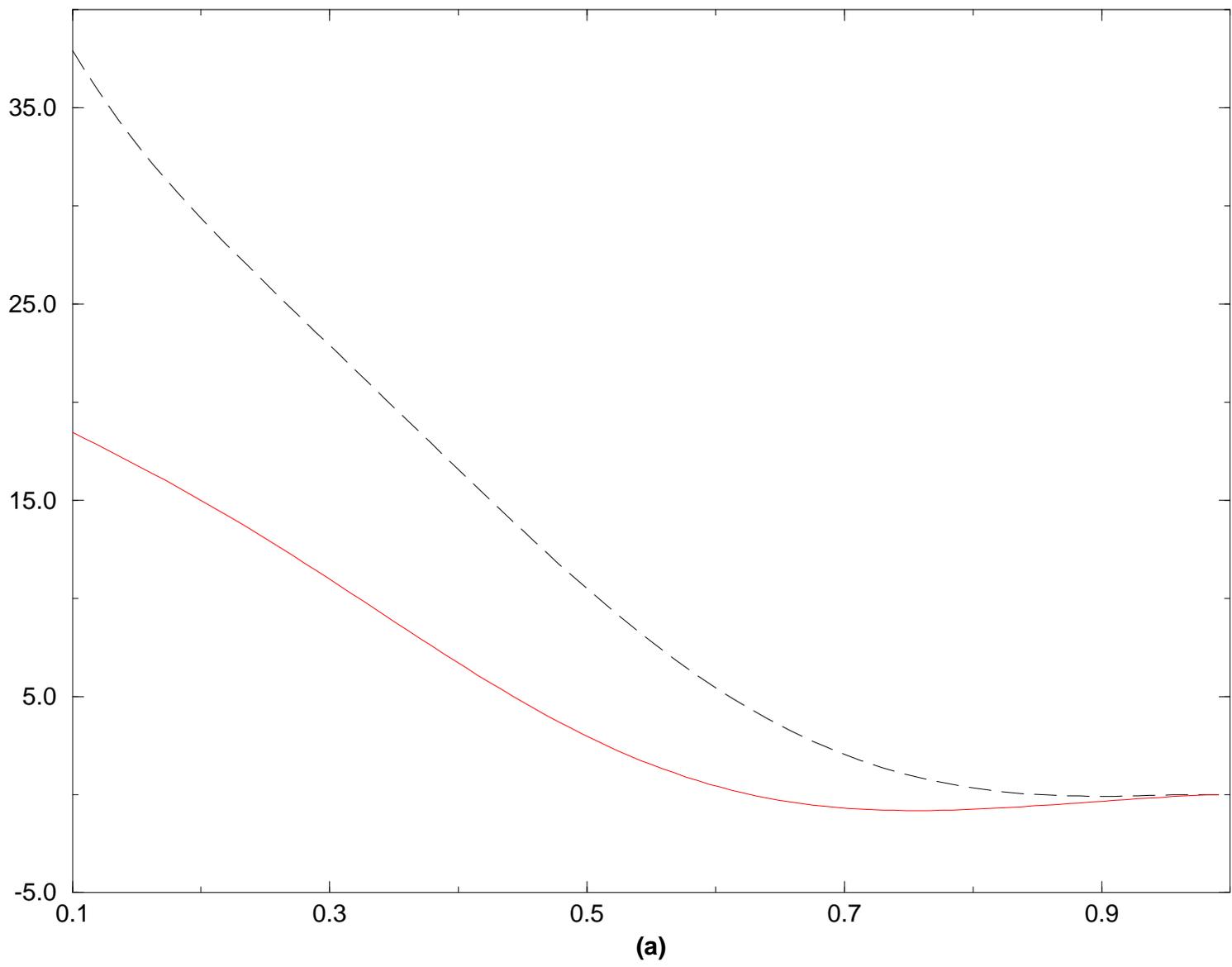
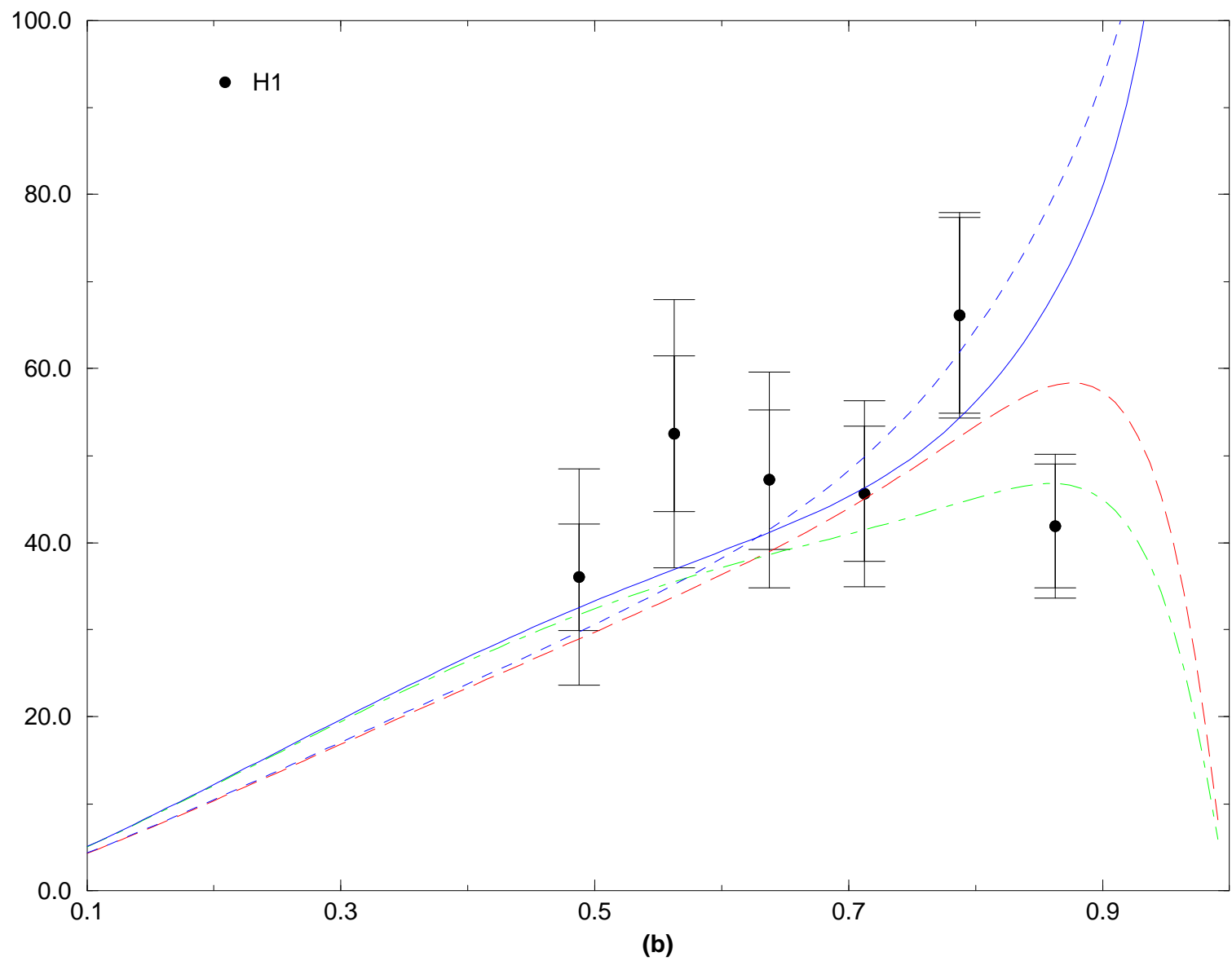
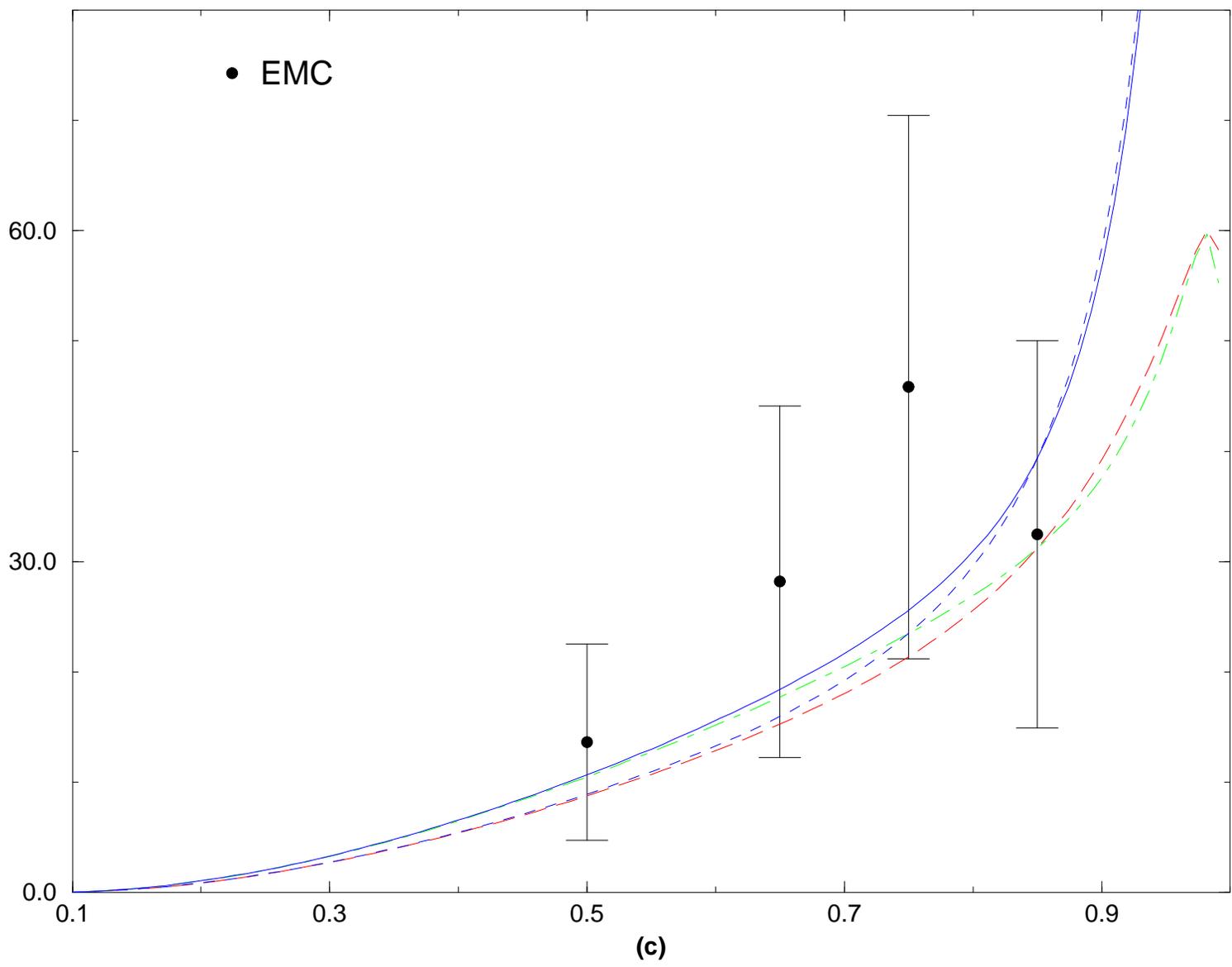
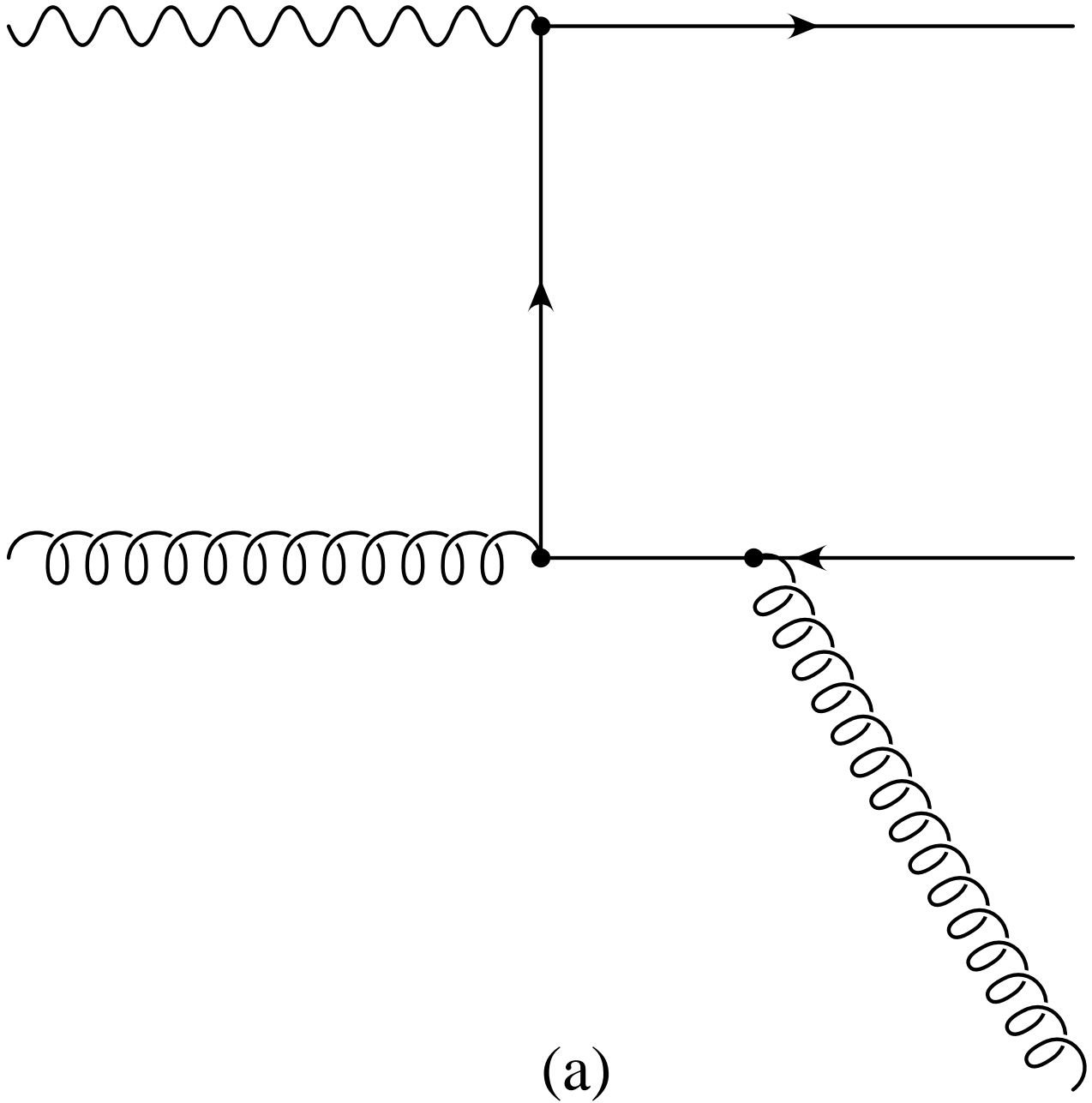
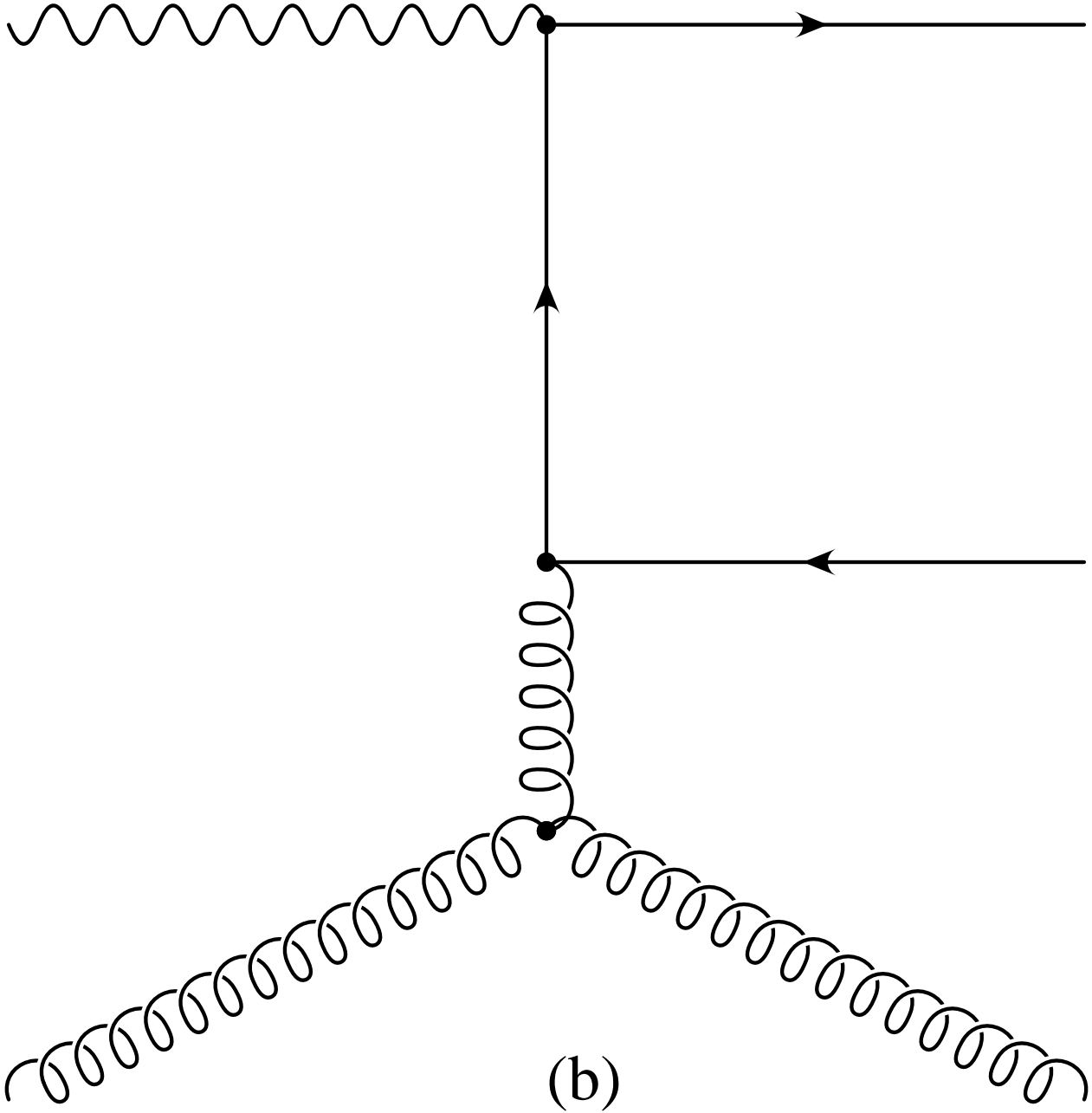
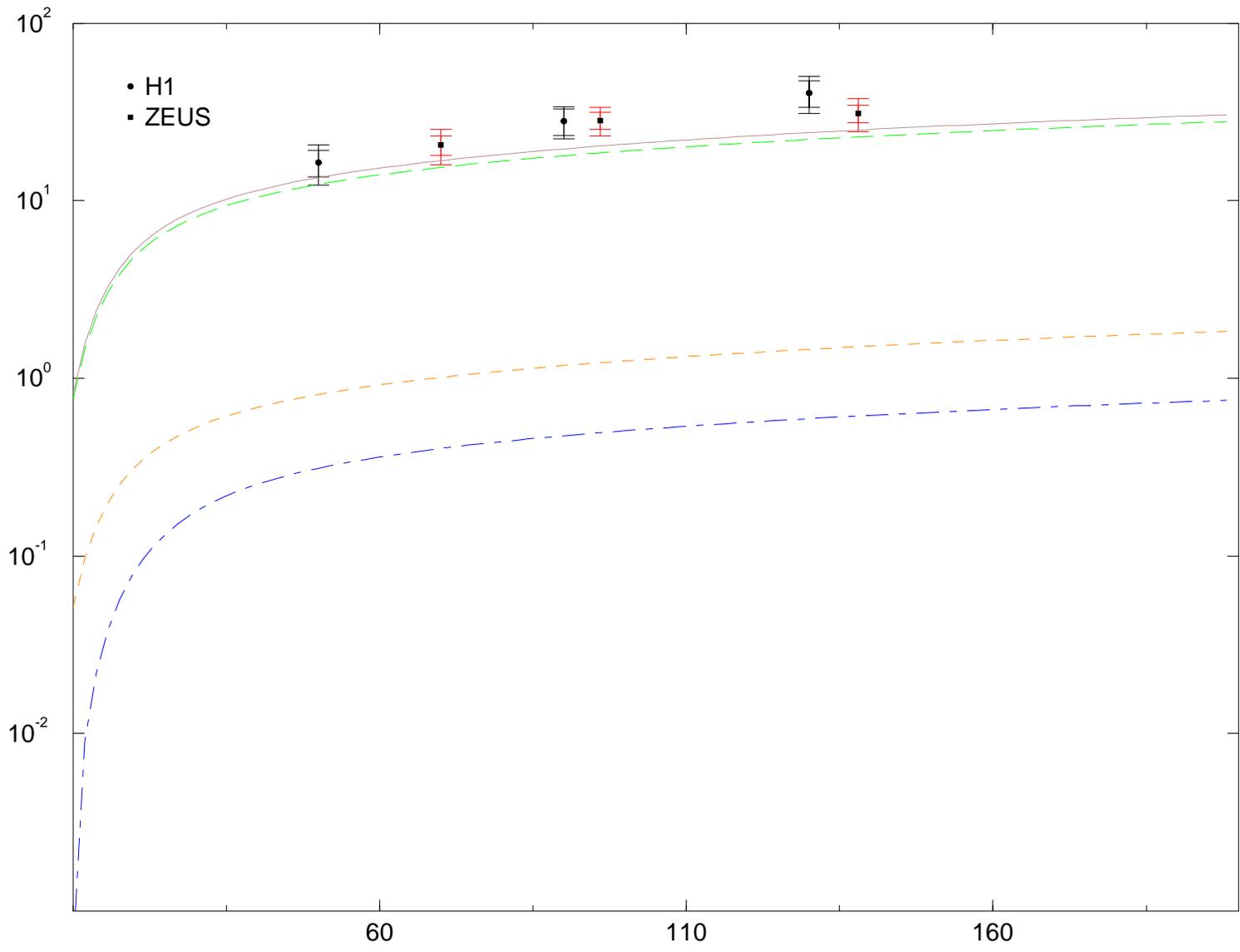
\end{document}